\documentclass[a4paper,12pt]{article}
 \usepackage{epsfig,float,wrapfig,amsmath,graphicx}
 \usepackage[hang,small,ruled,bf]{caption}
\newcommand{\be}{\begin{equation}}
\newcommand{\ee}{\end{equation}}
\newcommand{\beq}{\begin{eqnarray}}
\newcommand{\eeq}{\end{eqnarray}}

\begin{document}

{\Large
\begin{center}
{\bf Contribution of the nucleon-hyperon reaction channels to
K$^-$ production in proton-nucleus collisions}
\end{center} 
}
\begin{center}
{ H.W. Barz and L. Naumann\\
Forschungszentrum Rossendorf, 
Pf 510119, 01314 Dresden, Germany\\}

\vspace*{10mm}
%version 12 \today\\
\end{center}
\vspace*{10mm}
{\bf Abstract:}

The cross section for producing K$^-$ mesons in 
nucleon-hyperon collisions is estimated using the 
experimentally known pion-hyperon cross sections.
The results are implemented in a transport model which is 
applied to calculation of proton-nucleus collisions. 
Contrarily to earlier estimates in heavy-ion collisions 
the inclusion of the nucleon-hyperon 
cross section roughly doubles the K$^-$ production
in near-threshold proton-nucleus collisions.\\



%\newpage
%%%%%%%%%%%%%%%%%%%%%%%%%%%%%%%%%%%%%%%%%%%%%%%%%%%%%%%%%%%%%%%%%%%%%%%%

\section{Introduction}

The properties of strange mesons   within nuclear matter has 
been subject to numerous investigations.  Especially nuclear collisions
at energies near or below the production threshold of nucleon-nucleon
collisions should be very sensitive to the kaon properties in matter
\cite{Aichelin_Ko}. Early theoretical approaches 
based on effective chiral Lagrangians \cite{nelson} 
predicted an attractive
scalar potential which together with an isovector potential 
leads to a strong attractive K$^-$  and a moderately repulsive
K$^+$ potential. These potentials depend only weakly on the  momentum.
Indeed kaonic atoms require a strong attractive K$^-$ potential
and also the large K$^-$ rates observed 
in heavy-ion collisions,  
carried out by the FOPI \cite{FOPI1, FOPI2}
and KaoS \cite{KAOS1, KAOS2, KAOS3, KAOS4, KAOS5} collaborations, 
seemed to support these predictions. 
However,  more sophisticated 
theoretical investigations \cite{sibirtsev, schaffner,lutz} 
demonstrated a strong momentum dependence of the potentials 
which even became repulsive 
at large density and momentum for both K$^+$ and K$^-$ mesons.
These potentials do not comply with the early analyses 
of the measurements
of K$^-$ production. However one has to keep in mind that 
the elementary cross sections for K$^-$ production used are
not very well known from experiment. 

The knowledge of the production mechanism of kaons is a necessary 
condition to probe the theoretical predictions for the potentials.
In this respect it is useful to study also nucleon-nucleus collisions
as a further source of information.

In nucleus-nucleus collision antikaons have been observed 
at ion bombarding energy of 1.5~GeV per nucleon 
\cite{KAOS1, KAOS2, KAOS3, KAOS4, KAOS5, BEVALAC, GSI1}. 
This energy is far below the threshold energy of 2.5 GeV for nucleon-nucleon
collisions. This nucleon-nucleon production channel is unimportant 
in heavy-ion collisions 
due to the smallness of the kaon pair production cross section \cite{COSY11}
even if Fermi motion would help to overcome the threshold.
The antikaons can however be produced by 
multistep-scattering processes. 
In the sequence of the collisions combined with Fermi motion  
strangeness transfer via $\pi$Y $\to$ NK$^-$ 
reactions can take place with their comparably large cross sections.

Only a few data for antikaon production 
in nucleon-nucleus collision near threshold are available.
We compare our calculations with the KaoS data \cite{KAOS6}. 
A comparison to additional
existing antikaon data from FHS \cite{FHS2}  below 
and KEK-PS \cite{Sug98} above threshold seems not to be reasonable,
because of their very strong kinematical constraint.
In nucleon-nucleus collisions 
the above mentioned $\pi$Y channel is a rather improbable three step
process, because the single incoming proton alone has to produce 
both reaction partners before.
Also the second chance collision 
$\pi$N $\to$ NK$^+$K$^-$ channel has the small cross section
of the kaon pair production.
Therefore, in a second chance collision the antikaon 
could mainly be produced via the 
NY $\to$ NNK$^-$ channel 
which is open for an incident proton energy larger than 1.73 GeV.
Additional Fermi motion 
may  lead to a  further reduction of the threshold.  

Therefore we draw our attention to the  nucleon-hyperon
channel NY $\to$ NNK$^-$ cross section. This cross section
was already calculated early in the one-pion exchange model \cite{ko1}
and later in one-boson exchange approximation \cite{cassing1,cassing2} 
and was found to be unimportant for heavy-ion collisions.
Here we reevaluate the cross section within a different approach
avoiding the uncertainties arising from the badly known formfactors
when applying an effective perturbation theory. Our results
roughly agree with the cross sections obtained in 
ref.~\cite{cassing1,cassing3}.  In the relevant kinetic 
energy region the
cross section reaches nearly one mb and is 
about 50 times larger than the related cross sections
of the reactions NN$\to$ NYK$^+$. 
As the ratio of hyperons as well as of kaons to
the participating  nucleons is about 10$^{-4}$ this would lead
to a K$^-$ to K$^+$ ratio of about 10$^{-2}$ which is nearly  the 
magnitude of 
the measured ratio at 2.5 GeV beam energy. Thus, this channel 
should compete with the NN and $\pi$Y production channels.

In addition we mention that 
the in-medium cross sections may considerably differ from their vacuum values.
This was pointed out e.g. in ref. \cite{schaffner} where a 
considerable enhancement of the pion-hyperon channels have been
predicted. These results are based on
coupled channel calculations and  
are connected with a  shift  of the masses
of the $\Lambda(1405)$
and $\Sigma(1385)$ resonances in nuclear matter. 
This effect will not be considered here.

\section{Elementary cross sections} %%%%%%%%%%%%%%%%%%%%%%%%%%%%%%%%%%%

\begin{figure} 
\begin{center}
\includegraphics[width=100mm]{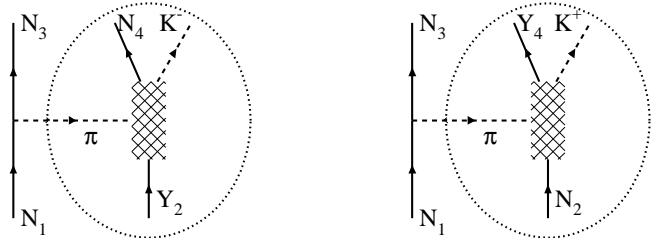}
\end{center}
\caption{Diagram for antikaon (left) and kaon (right)
production in the
process N+Y(N) $\rightarrow$ N+N(Y)+K$^-$(K$^+$) by pion exchange.
The $T$ matrix describing processes within 
the hatched box is determined from experimental data.}
\label{diag}
\end{figure}

To estimate the  NY $\to$ NNK$^-$  cross sections we start with 
the Feynman diagrams shown in Fig.~\ref{diag} which are similar 
for NY $\to$ NNK$^-$ and NN $\to$ NYK$^+$ processes.
The K$^\pm$ mesons are generated by the subprocess where the
intermediate meson (here a pion) 
interacts with the second baryon. 
This subprocess cannot be calculated  for certainty
because many resonances contribute with unknown
coupling constants. The results of such one-boson exchange reactions
can be found in ref. \cite{cassing1}. 
Here we will approach the study of these cross sections differently 
by making the assumption
that the most important meson exchange is that of a pion. The $\pi$B $\to$ K$^\pm$ 
cross sections are known experimentally from which we can extract the
square of the transition matrix elements 
$T_{\pi {\rm B}}$ illustrated by the hatched areas
in Fig.~\ref{diag}. Then, these values are
used to calculate the cross sections in accordance with 
the diagram in Fig.~\ref{diag}.
We calculate both the kaon and the antikaon 
production in order to check the method since the 
cross sections for NN $\to$ NYK$^+$ \cite{tsushima1} has been calculated and 
adjusted to the partially known pp cross sections \cite{landolt}.

Thus we consider the process N$_1$+B$_2$ $\to$ N$_3$+B$_4$ + 
K($\bar {\rm K}$), 
where the symbol N
denotes a nucleon with isospin $I_1=I_3=1/2$ and one of the symbols
B$_2$ or B$_4$ 
stands for a hyperon.
The pion-nucleon coupling in the left hand vertex in Fig.~\ref{diag} 
is described by the lagrangian
\be
{\cal L}_{\pi NN} = g \bar {\psi}\gamma^5 \vec{\tau} \psi \vec{\pi}
\ee
with $\psi$ denoting the nucleon field, $\vec{\pi}$ the 
pion field,  $\vec{\tau}$ the isospin Pauli matrix, $\gamma^5$ is a
Dirac matrix, and g = 13.6 fixes the pion-nucleon
coupling constant.
The spin and isospin averaged 
cross section for the process 
at the center-of-mass energy $\sqrt{s}$ reads
\beq
\label{defsigma}
\nonumber
\sigma_{\rm K} &=&  \frac{g^2}{2\lambda^{1/2}(m_1^2, m_2^2, s)} 
         \frac{1}{(2\pi)^5}  \frac{1}{8(2I_2+1)}   \\[1mm]
      &\times& 
	   \sum_{spin,isospin}  |<I_3|\vec{\tau}|I_1>|^2
        \int  d^4p_\pi \frac{d^3{\bf p}_4}{2 p_4^0} 
         \frac{d^3{\bf p}_K}{2 p_K^0}  
	  \mid (\bar{u}_3 \gamma^5 u_1) \mid^2\\
     & \times &  \nonumber
         \delta((p_1-p_\pi)^2-m_3^2) 
         \delta^4(p_\pi+p_2 - p_3 -p_K)    
	   \mid  \frac{1}{p_\pi^2-m_\pi^2}
	            T_{\pi B_2: B_4K} \mid^2,
\eeq
where the indices of the momenta $p$ and masses $m$ are those
used in Fig.~\ref{diag}, $u$ denotes the nucleon spinor, and
$\lambda(a,b,c) = (s-a-b)^2 - (2ab)^2$ is the triangle function,
and $I_2$ stands for the isospin of particle B$_2$.
The integral over the outgoing momentum $p_3$ of particle N$_3$
has been substituted by the pion momentum $p_\pi=p_1-p_3$. 
The symbol $ T_{\pi B_2: B_4K}$ represents the 
encircled part in Fig.~\ref{diag} which determines the kaon (antikaon)
production in pion-baryon collisions the cross section of which is
given by 

\beq
\label{subsigma}
\nonumber
\sigma_{\pi B_2} &=&  
         \frac{1}{2\lambda^{1/2}(m_\pi^2, m_2^2, s_{\pi B_2})}
         \frac{1}{(2\pi)^2}  \frac{1}{6(2I_2+1)} \\[1mm]
      &\times& 
	   \sum_{spin,isospin} 
        \int  \frac{d^3{\bf p}_4}{2 p_4^0} 
         \frac{d^3{\bf p}_K}{2 p_K^0}  
         \delta^4(p_\pi+p_2 - p_3 -p_K)   
         \mid T_{\pi B_2: B_4K} \mid^2. 
\eeq
This cross section depends on the square of the center-of-mass energy 
$s_{\pi B_2} = (p_\pi + p_2)^2 $ with the on-shell condition
$p^0_\pi = \sqrt{m_\pi^2+{\bf p}_\pi^2}$. Notice that 
in Eq.(\ref{defsigma}) the pion momentum $p_\pi$ is off-shell.

\begin{figure}
\begin{center}
\includegraphics[width=86mm]{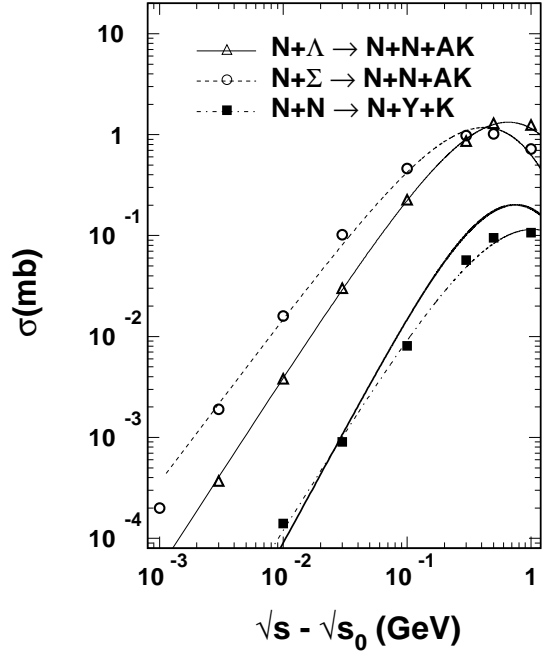}
\end{center}
\caption{Isospin averaged production cross sections for
kaon (full squares) and antikaon (open symbols)
production versus excess energy. The thin lines are parametrizations
(described in text) adjusted
to the calculated values (symbols) whereas the thick line
displays the cross section for K$^+$ production
as calculated in ref. \cite{tsushima1}.}
\label{crosssect}
\end{figure}

Inserting Eq.(\ref{subsigma}) into Eq.(\ref{defsigma}),  
summing over the spin quantum numbers of the nucleons and
integrating out the time component $p_\pi^0$ of the pion momentum
we obtain the cross section 
\beq
\sigma_{\rm K} &=&  \nonumber  \label{final}
         \frac{g^2}{\lambda^{1/2}(m_1^2, m_2^2,  s)} 
         \frac{1}{(2\pi)^3} \int  \frac{d^3 {\bf p}_\pi}
	   {\sqrt{m_3^2+{\bf p}^2_\pi}}
	     \frac{p_1\cdot p_3 - m_1m_3}{(p_\pi^2-m_\pi^2)^2}\\[1mm]
     &\times &  f^4(p_\pi) \lambda^{1/2}(m_\pi^2, m_2^2,  s_{\pi B_2})
       \sum_{I_\pi} \sigma_{\pi B_2}(s_{\pi B_2}), 
\eeq
where the formfactor 
\be  \label{formf}
f(p_\pi) \,=\, \frac{\Lambda^2-m_\pi^2}{\Lambda^2-p_\pi^2}\, 
\ee
with $\Lambda$ = 1.6 GeV \cite{machleidt} has been introduced.

\begin{figure}
\begin{center}
\includegraphics[width=86mm]{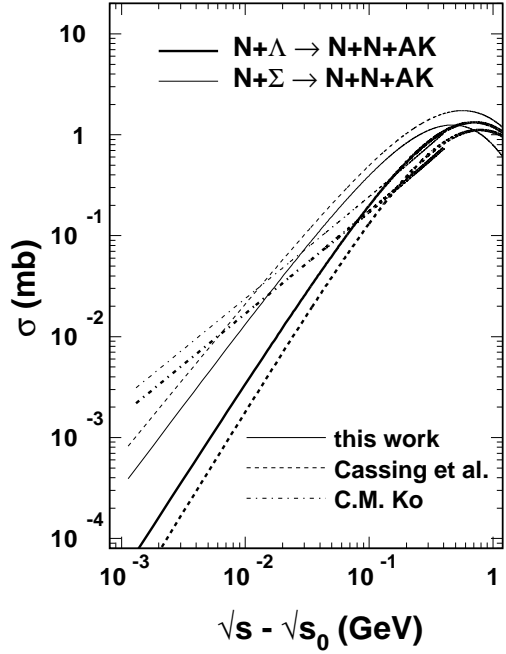}
\end{center}
\caption{Comparison of isospin averaged production cross 
sections for
antikaon production versus excess energy  obtained by various
models. The thick (thin) curves describe the N$\Lambda$ (N$\Sigma$)
collisons.  The full lines depict our results, the dot-dashed and dashed
lines are results from \cite{ko1} and \cite{cassing3}, respectively.}
\label{comparison}
\end{figure}

The cross sections  $\sigma_{\pi B_2}$ needed in Eq.(\ref{final}) for the 
reactions $\pi$Y $\to$ NK$^-$ can be derived
from the measured inverse reactions K$^-$p $\to$ $\Sigma^+\pi^-$, 
$\Lambda\pi^0$, $\Sigma^-\pi^+$ and  K$^-$n $\to$ 
$\Lambda\pi^-$, $\Sigma^-\pi^0$ 
which are given in ref. \cite{landolt}. The
K$^0$N cross sections are derived by  isospin reflection.

The results for the antikaon production in 
N$\Sigma$ and N$\Lambda$ collisions
are represented by the open symbols in Fig. \ref{crosssect}.
The lines through the symbols show fits with the standard 
parametrization $\sigma \propto (s-s_0)^a/s^b $.
The K$^+$ cross section is compared to the parametrization of
ref. \cite{tsushima1} which gives in the low energy
region nearly the same cross section  but underestimates
the cross section above 100 MeV.
The parametrization in ref. \cite{tsushima1}
is based on a model which uses diagrams
which have the same structure as those of Fig. \ref{diag}.
In that  investigation  an exchange of $\pi$, $\eta$ or  $\rho$
mesons is included, and it is shown that the pion gives
the main contribution, a fact that supports our assumption
of the dominance of the pion exchange.

In Fig.~\ref{comparison} we compare the curves in Fig.~\ref{crosssect}
to the results of previous calculations \cite{ko1,cassing1,cassing3}.
There are only small deviations from the calculations
in the one-boson exchange approximation \cite{cassing1,cassing3} which
also shows the dominance of the N$\Sigma$ channel. The 
result  \cite{ko1} has  a weaker energy dependence but provides 
comparable cross section in the relevant energy region of
about 100 MeV above threshold. In this investigation it was also
shown that this channel could contribute
about 10\% to the K$^-$ production in heavy-ion collisions.
In our following calculations we find that even half of the antikaons
can stem from the NY channel in proton-nucleus collisions.

\section{Comparison to data }

It is our aim to study the role of the NY $\to$ NNK$^-$ reaction
in pA collisions. Usually these channels are not included in 
standard analyses of those reactions which consider only the elementary
 BB and $\pi$B collisions. To study this question
we additionally incorporate the \mbox{NY channels} 
into a transport model calculation which is based
on the Boltzmann-\"Uhling-Uhlenbeck equation \cite{wolf1}.
Furthermore, the production rate of the K$^-$ mesons also depends 
sensitively on the attractive K$^-$ potential. Therefore, the
inclusion of different production channels will effect 
predictions on the size of this potential when derived from comparison
with data.

\begin{figure} 
\begin{center}
\includegraphics[height=100mm]{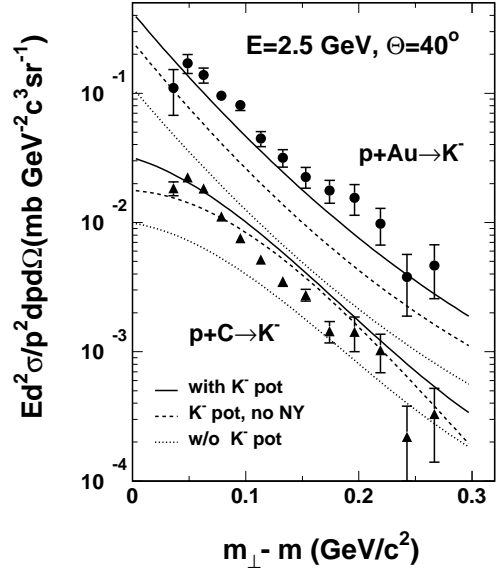}
\end{center}
\caption{ Comparison of measured \cite{KAOS6} invariant 
K$^-$ meson cross sections (symbols) as a function of the
transverse mass with calculations for 
proton collisions on C and Au targets at 2.5 GeV beam energy. 
The full (dotted) lines are calculated with (without)
an antikaon potential while the dashed curves are obtained
without 
the contribution of the nucleon-hyperon channel. 
}
\label{parvar}
\end{figure}

In Fig. \ref{parvar} we present the differential cross sections
at a laboratory angle of 40$^\circ$ for collisions of protons with 
$^{12}$C and $^{197}$Au at 2.5 GeV beam energy. 
The dotted lines are calculated without using potentials for the
kaons and antikaons. These calculations 
underestimate clearly the measured data \cite{KAOS6}. 
The attractive antikaon potential,
\beq 
V_{\bar{K}}\,=\,-0.08\,GeV \, \frac{n}{n_0}\,
\eeq
in addition with the \mbox{NY channels} leads to an increase of the 
cross section as shown by the full lines. Such a potential improves 
the agreement with the data.
The dashed curves are results where the NY$\to$K$^-$ channels 
has been excluded.
Disregarding these channels 
the cross section diminishes by about 50\% in p+Au collisions. 
This shows the importance of the NY$\to$K$^-$ channels when
one intends to determine the K$^-$ potential.
For the 
light C target
the influence is much smaller as the hyperons have a smaller 
chance to collide with further nucleons before leaving the reaction zone.\\

\begin{figure} 
\begin{center}
\includegraphics[width=140mm]{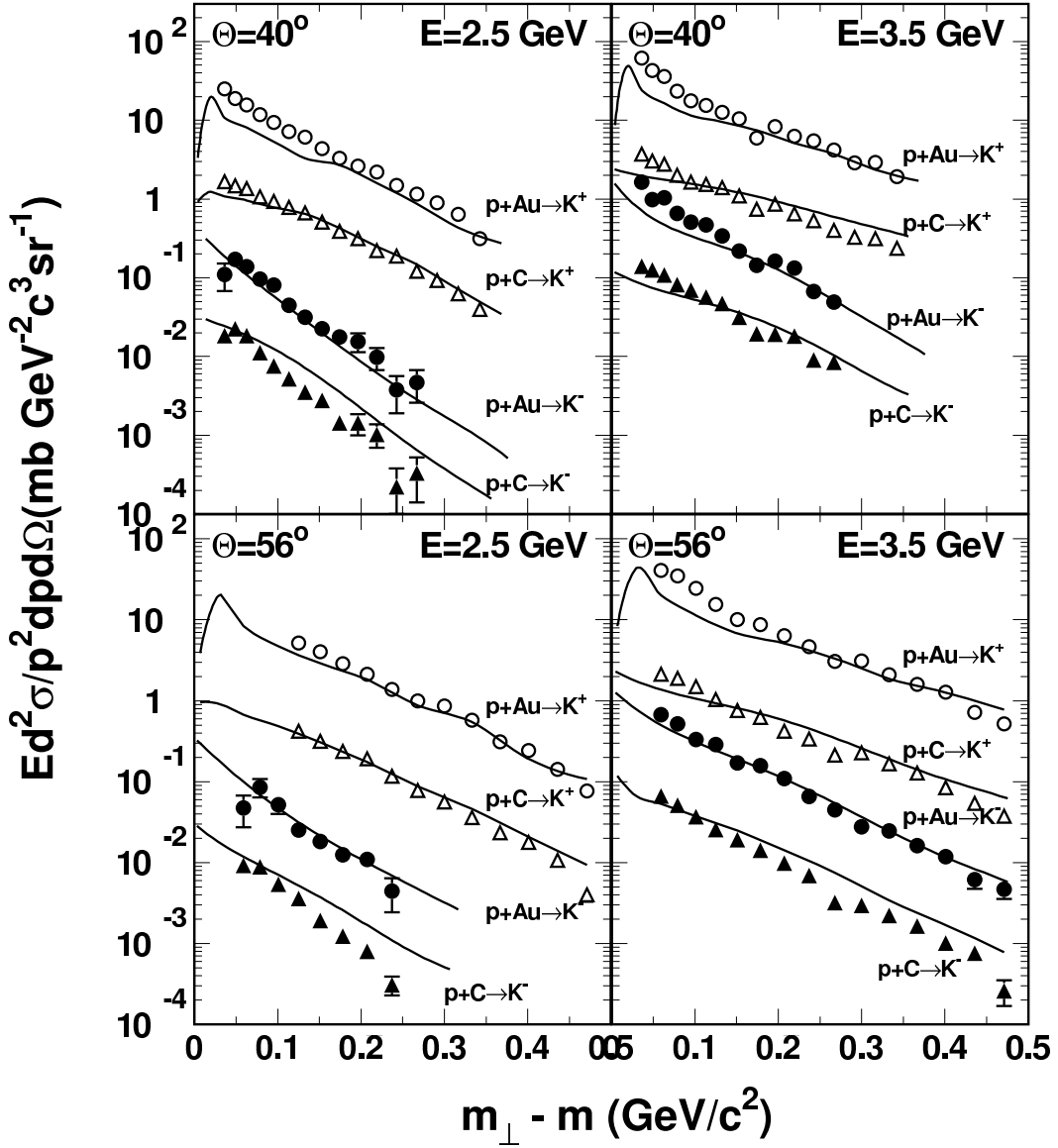}
\end{center}
\caption{ Invariant differential K$^\pm$ cross section versus  K-meson 
transverse mass at 2.5 and 3.5 GeV proton beam energy. 
The solid lines refer to our
calculations including the NY channels for the antikaon production.
Data are taken from ref.~\cite{KAOS6}.
}
\label{compar}
\end{figure}

Finally, we compare in Fig.~\ref{compar} our calculations 
for K$^+$ and  K$^-$ production 
with data obtained by the KaoS collaboration \cite{KAOS6}
for proton-nucleus collisions at bombarding energies of 2.5 GeV and  
3.5 GeV on C and Au targets and K meson emission angles of 
$40^{\circ}$ and $56^{\circ}$.
The kinetic beam energy ${T_\mathrm{kin}=2.5}$~GeV was close to the
production threshold in nucleon-nucleon
collisions.
In the calculations we have used the parametrizations of
ref.~\cite{tsushima1,tsushima2} and
obtained kaon cross sections which are slightly smaller 
than the data for the gold target  for both angles.

The kaon and antikaon production yields are related in the 
threshold region since nearly all of the antikaons are created in
collisions of pions and nucleons with hyperons, the number
of which equals to the kaon number because of strangeness
conservation.
This fact also holds if a chemical equilibrium
between K$^-$ and Y is reached via the 
\mbox{K$^-$N $\leftrightarrow$ $\pi$Y} reaction \cite{librown,aichelin}.
Therefore it is interesting to compare
the ratio of the angle integrated cross sections
for K$^-$ to those of K$^+$. In Tab. 1 we display the calculated values
and the experimental results obtained  
within the measured phase space  by the KaoS collaboration \cite{KAOS6}. 
Our calculations overestimate this ratio for 2.5 GeV but underestimate
it for the higher energy of 3.5 GeV.

{\sl\small{\bf Tab.\,1}
  Comparison of calculated and experimentally  obtained antikaon  
  to kaon cross-section ratios  
  for proton collisions on carbon $R(C)$ and gold $R(Au)$.
  Experimental values with errors of about 20\%
  are taken from \cite{KAOS6}.\\}
\vskip3mm
\begin{tabular}{|c||c|c|c|c|}
\hline
$T_{kin}$(GeV) &$R_{exp}(C)$&$R_{calc}(C)$&$R_{exp}(Au)$&$R_{calc}(Au)$ \\
\hline
2.5& 0.0085 & 0.013  & 0.0074  & 0.010 \\
\hline
3.5& 0.028  & 0.024 & 0.027   & 0.024  \\
\hline
\end{tabular}

\section{Conclusions}

We have calculated the
cross sections of  antikaon-production in 
near-threshold proton collisions on carbon and gold targets. 
Comparison of our calculations with both kaon and antikaon data
from  the KaoS collaboration 
\cite{KAOS6} were made at beam energies of 2.5 and 3.5 GeV  for
laboratory angles of $40^{\circ}$ and $56^{\circ}$.      
Including the  NY $\to$ NNK$^-$ channels nearly doubles the antikaon
yield for collisions of protons on heavy targets like Au.
No significant influence was found for the light carbon target.
Calculated ratios of K$^-$ to K$^+$ cross sections
came also reasonably close to the data.  
We conclude that the  NY $\to$ NNK$^-$ channels 
have to be included in realistic
calculations, especially for heavy targets, 
 in order to study the properties of strange mesons                 
in nuclear matter at normal nuclear density.

\subsection*{Acknowledgments} %%%%%%%%%%%%%%%%%%%%%%%%%%%%%%%%%%%%%

Valuable discussions with Werner Scheinast and Burkhard 
K\"ampfer are acknowledged.
We thank Christoph  Hartnack for informing us on similar results in
his study.
This work was supported in part by the BMBF grant 06DR121.

\end{document}